\documentclass[conference]{IEEEtran}
\IEEEoverridecommandlockouts
\usepackage{amsmath,amssymb,amsfonts}
\usepackage{algorithmic}
\usepackage{graphicx}
\usepackage{multirow}
\usepackage{textcomp}
\usepackage{xcolor}
\usepackage{subcaption}
\def\BibTeX{{\rm B\kern-.05em{\sc i\kern-.025em b}\kern-.08em
    T\kern-.1667em\lower.7ex\hbox{E}\kern-.125emX}}
\usepackage[backend=biber]{biblatex}
\begin{document}

\title{From Design to Validation: Preparing a LEO-Capable UE for End-to-End System Evaluation\\
\thanks{This work has been supported by the project TRANTOR which has received funding from the European Union’s Horizon Europe research and innovation program under grant agreement No. 101081983 and by MINECO and EU – NextGenerationEU [UNICO-5G I+D/AROMA3D-SPACE (TSI-063000-2021-70)].}
}

\author{\IEEEauthorblockN{Amedeo Giuliani, Pol Henarejos,\\Erislandy Mozo, and M\`{a}rius Caus}
\IEEEauthorblockA{\textit{Centre Tecnol\`{o}gic de Telecomunicacions}\\
\textit{de Catalunya (CTTC)}\\
Castelldefels, Barcelona, Spain\\
\{amedeo.giuliani,pol.henarejos,\\erislandy.mozo,marius.caus\}@cttc.cat}
\and
\IEEEauthorblockN{Miguel \'{A}ngel Solis Gallego\\and Jaime Su\'{a}rez Garc\'{i}a}
\IEEEauthorblockA{\textit{Inster Grupo Oesia}\\
Nigr\'{a}n, Pontevedra, Spain\\
\{masolis,jsuarez\}@oesia.com}
\and
\IEEEauthorblockN{Rami Othman}
\IEEEauthorblockA{\textit{QuadSAT}\\
Odense, Denmark\\
ro@quadsat.com}
\and
\IEEEauthorblockN{Justin Tallon}
\IEEEauthorblockA{\textit{Software Radio System (SRS)}\\
Barcelona, Spain\\
justin.tallon@srs.io}
}
\maketitle

\begin{abstract}
The extension of 5G connectivity through Low-Earth Orbit satellite systems introduces significant technical challenges, particularly due to time-varying propagation delays and high Doppler shifts resulting from satellite motion. While the Third Generation Partnership Project Release 17 established the initial framework for non-terrestrial networks, the ongoing developments in Release 19 further enhance this effort by introducing support for regenerative payload architectures, where part of the communication protocol stack is processed directly on board the satellite. In this work, we present the design of a 5G user equipment adapted for Low-Earth Orbit satellite connectivity, with specific focus on strategies for managing variable delay and Doppler compensation. Additionally, we describe a custom experimental platform based on a drone-mounted software-defined radio platform capable of emulating both transparent and regenerative satellite payloads. Although full end-to-end system validation is not yet complete, initial laboratory tests confirm the feasibility of the architecture and lay the groundwork for future experimental campaigns.
\end{abstract}


\section{Introduction}
The integration of Low-Earth Orbit (LEO) satellite systems into 5G networks has emerged as a critical area of research to ensure global connectivity and extend cellular services to remote or underserved regions. The 3rd Generation Partnership Project (3GPP) has addressed these challenges through the standardization efforts of 5G Non-Terrestrial Networks (NTN), especially under Release 17 \cite{darwish2022leo}\cite{lin20215g}. Recent developments in 3GPP Release 19 have continued to expand the NTN framework, with particular attention given to regenerative payload architectures, in which satellite payloads perform partial or full protocol stack processing on board. This advancement enables reduced round-trip delay, enhanced spectral efficiency, and more flexible network deployments—especially relevant for LEO constellations. By supporting on-board new generation Node Base (gNB) functionality and facilitating direct inter-satellite connectivity, regenerative payloads are set to play a key role in future 5G-Advanced and 6G satellite systems. In this context, our work takes a step toward enabling such architectures by emulating a regenerative scenario using a drone-based platform equipped with a digital payload.

While most prior research has concentrated on analytical and simulation-based evaluations, real-world experimentation remains limited—particularly involving actual or emulated LEO conditions. This work presents a practical approach to designing and validating a 5G user equipment (UE) capable of supporting LEO-based NTN connectivity. Building on previous efforts focused on geostationary satellite communication, we extend the system design to account for the unique challenges posed by LEO orbits, such as time-varying propagation delays and high Doppler shifts. Moreover, this paper details the modifications to the UE architecture, synchronization procedures, and compensation strategies, as well as the field setup using a drone-based satellite emulator, to conduct end-to-end system-level testing under realistic LEO conditions. 
\section{Previous Work}
Very few works about testing 5G NR NTN LEO settings have been published. Most of them regard analytical and simulation results. For example, in \cite{sedin2020throughput} authors studied and evaluated the throughput and capacity of LEO 5G NR NTN. In detail, they carried out system simulations based on the latest developments from 3GPP including channel models, antenna models, and network scenarios, to calculate the achievable throughput. They also derived analytical expressions for LEO NTN system capacity estimation and estimated the system capacity of a mega-LEO constellation.

The project 5G-LEO \cite{5g-leo} focuses on extending the Open Air Interface (OAI) suite \cite{oai-project} (composed of a UE and a gNB) toward the 3GPP Release 17. The progress is detailed in \cite{kumar20235g}, in which the authors presented the modifications that must be performed on the OAI stack in order for the UE to handle time-varying delay and high Doppler, typical of LEO settings, and successfully connecting to the network.
%But at the time of such work, the OAI suite was lacking the implementation of the System Information Block (SIB) 19, that is, the development was not fully compliant of Release 17. The info contained in the SIB 19 is likely to have been supplied to the UE at execution time via a configuration file.

The present work directly builds up onto our previous work regarding applying Rel-17 modifications on the UE side to enable 5G NR NTN connectivity through geostationary orbit (GSO) satellites, in which we set up a testbed with composed of a real GSO satellite operating at Ku band and two dish antennas that the developed UE and the gNB use to communicate to the satellite in the sky \cite{giuliani2025}.
\section{System Model}
\begin{figure}
    \centering
    \includegraphics[width=\linewidth]{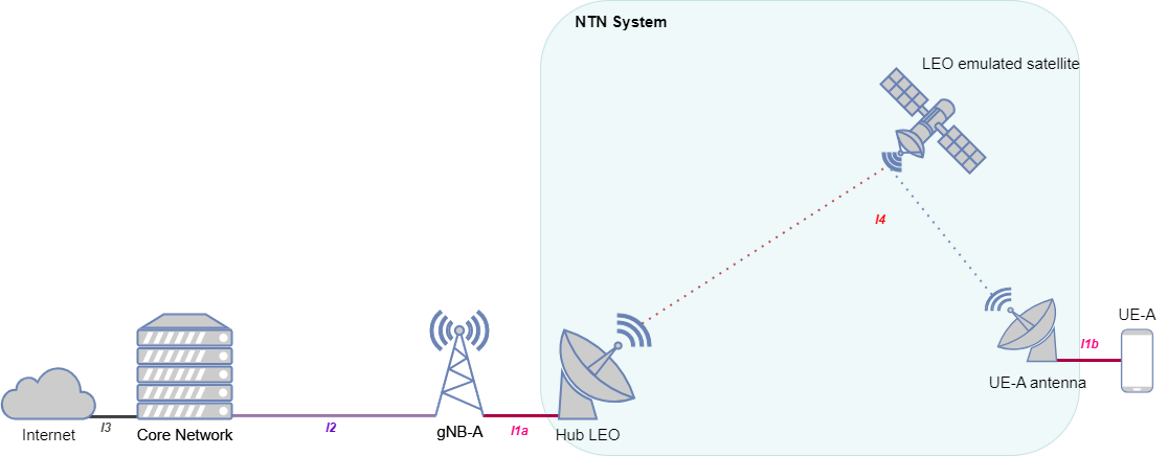}
    \caption{System model.}
    \label{fig:system-model}
\end{figure}

The architecture of this work is the following. On the ground, there are the UE and the gNB, both equipped with an antenna working at Ka band, in order to connect to the (emulated) LEO satellite in the sky. The gNB is connected to the standalone (SA) 5G core network (CN) which is in turn connected to the data network to provide internet access to the UE. This is illustrated in Fig. \ref{fig:system-model}.

The gNB is provided by SRS under the srsRAN Project \cite{srsran-project}. It supports both FDD and TDD, all FR1 bands, 15 kHz and 30 kHz SCS, up to 256 Quadrature Amplitude Modulation (QAM). It is Release 17 compliant, with non-terrestrial support for GSO/non-GSO satellites.

The CN used is provided by the open-source project open5Gs \cite{open5gs-project}, configured to work in 5G SA mode.
\section{UE Design}
The UE has been extended to support connectivity with LEO satellites. Unlike GSO satellites, which maintain a fixed position relative to the Earth, LEO satellites follow dynamic orbits that cause the distance between the satellite and the UE to vary continuously over time. Due to the high relative velocity of LEO satellites with respect to the Earth, significant Doppler shifts are introduced.

The distance profile between the UE and a LEO satellite during a visibility window typically follows a parabolic trajectory. The satellite reached its highest altitudes – the start and the end of the visibility window – at the two outer points of the curve. The lowest point of the curve, where the satellite is closest to the UE, occurs when the satellite is at 90º elevation with respect to the UE, that is, is directly above the UE position.

The Doppler shift, closely tied to this distance variation, follows a shape resembling an inverted sigmoid function. At the start of the visibility window, when the satellite is at its highest altitude and approaching the UE at maximum relative speed, the Doppler shift is at its positive peak. As the satellite moves towards the UE, the relative speed decreases and therefore also the Doppler shift starts to decrease. When the satellite reaches the point in which it has the minimum distance with respect to the UE (the point in which it is at 90º inclination with respect to the UE), the relative motion briefly ceases in the radial direction, and the Doppler shift momentarily drops to zero. Once the satellite passes this point and begins moving away from the UE, the Doppler shift starts increasing again in magnitude, but with a negative sign, reflecting the increasing satellite’s receding speed.
\subsection{Variable Delay}
A straightforward approach would be to just rely on Timing Advance (TA) commands issued by the gNB to the UE to address the varying propagation delay in LEO scenarios. However, this solution presents two major challenges:
\begin{itemize}
    \item There is no guarantee that the rate of the change in propagation delay is slower than the rate at which the gNB can send TA commands (and for the UE to apply them) without incurring in timing misalignments
    \item Adjusting the delay through TA commands, can result in negative timing values and although such values are technically allowed by the standard, they can lead to loss of samples, especially if applied during slots carrying uplink (UL) data, effectively causing the loss of the entire slot.
\end{itemize}
Therefore, the following strategy has been implemented. In UL, the UE pre-compensates for a fixed delay equal to the maximum expected round-trip delay within the NTN cell, denoted as $k_{offset}$. The UE will make use of a buffer to emulate such a delay at any time. Then, when the actual delay d is less than $k_{offset}$, the UE inserts an artificial delay by updating the buffer size by the difference $k_{offset}-d$, effectively delaying the UL packets by just the right quantity to match the delay the UE is pre-compensating for, allowing for proper synchronization at the gNB.

\begin{figure}
    \centering
    \includegraphics[width=\linewidth]{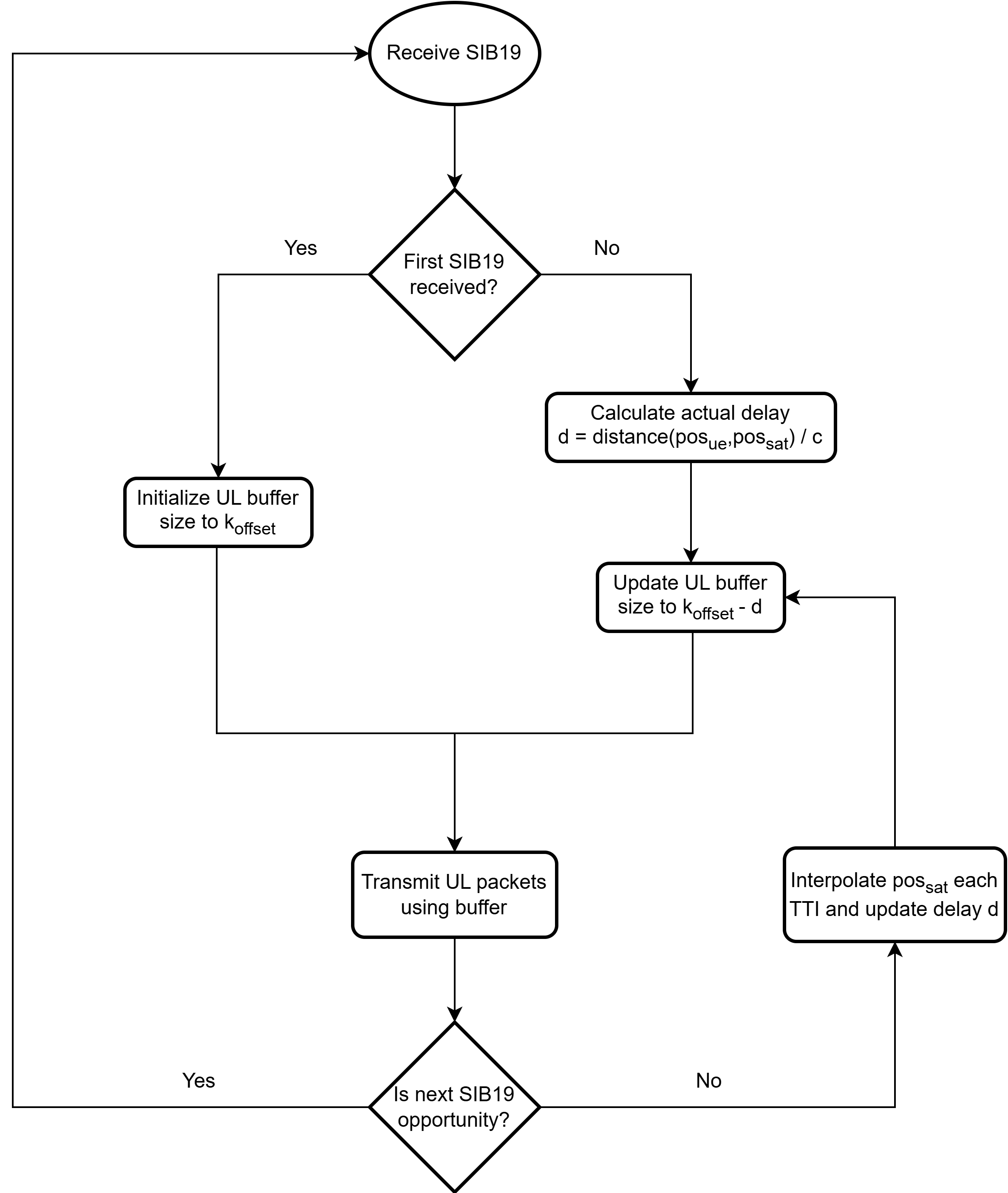}
    \caption{Flowchart of the delay compensation process performed by the UE.}
    \label{fig:delay-compensation-process}
\end{figure}

The flowchart in Figure \ref{fig:delay-compensation-process} resumes the procedure described above, that the UE follows each time that a System Information Block (SIB) 19 is received. Just for reference, the SIB 19 is a dedicated SIB for NTN introduced by 3GPP in Release 17 that contains crucial information about the NTN cell and the satellite.
\subsection{High Doppler}
\subsubsection{Downlink Compensation}
\begin{figure}
    \centering
    \includegraphics[width=\linewidth]{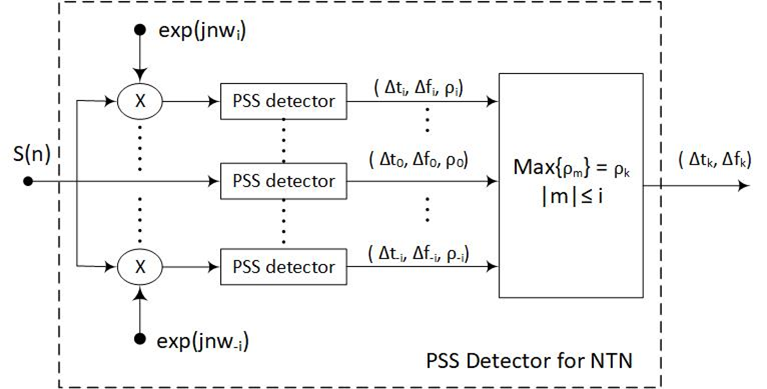}
    \caption{PSS detector for NTN.}
    \label{fig:pss-detector-ntn}
\end{figure}

The downlink (DL) synchronization procedure is performed by the UE to acquire the timing and the carrier frequency. Towards this end, the UE exploits the Synchronization Signal Block (SSB) that is broadcasted with a default periodicity of 20 ms. 
The Doppler frequency shift that stems from the orbital motion becomes the main impairment to locate the SSB in the DL frame. Provided that the resulting magnitude is lower than the subcarrier spacing of the SSB, the synchronization can be performed with the same algorithm used in terrestrial deployments. Otherwise, if the Carrier Frequency Offset (CFO) is larger than the subcarrier spacing, additional complexity is needed at the UE receiver to achieve initial DL synchronization.
The preferred solution increases the robustness of the CFO by resorting to parallel correlations. The idea is to perform a 2D search in time and frequency dimensions. To this end, the local sequence used in each correlator is matched to a different CFO candidate. The larger the CFO range, the more parallel correlators are needed. By selecting the branch that exhibits the highest correlation peak, the uncertainty range is reduced, and the probability of detection is improved. This concept is illustrated in \ref{fig:pss-detector-ntn}.

\subsubsection{Uplink Compensation}
Once the UE has estimated the DL Doppler shift caused by the satellite motion, it must also calculate and account for the corresponding UL Doppler shift to ensure successful packet transmission. These two values are time varying, and must be recalculated at specific time intervals, as detailed below.
Regarding the DL Doppler, it is updated every time an SSB is received, and the procedure described above is repeated each time, which also includes the Doppler component introduced by the local oscillator of the UE.

Instead, the UL Doppler is updated upon reception of each SIB 19. In particular, the SIB 19 contains, among other parameters, the DL and UL Doppler shifts resulting from the satellite movement. To obtain the complete UL Doppler shift, the UE must also account for the frequency offset due to its local oscillator. Since this contribution is already embedded in the DL Doppler estimate, the total UL Doppler shift can be derived using the following formula:
\begin{equation}
    f_{UL} = f_{DL}^{SSB} - f_{DL}^{SIB19} - f_{UL}^{SIB19}
\end{equation}

\begin{figure}
    \centering
    \includegraphics[width=\linewidth]{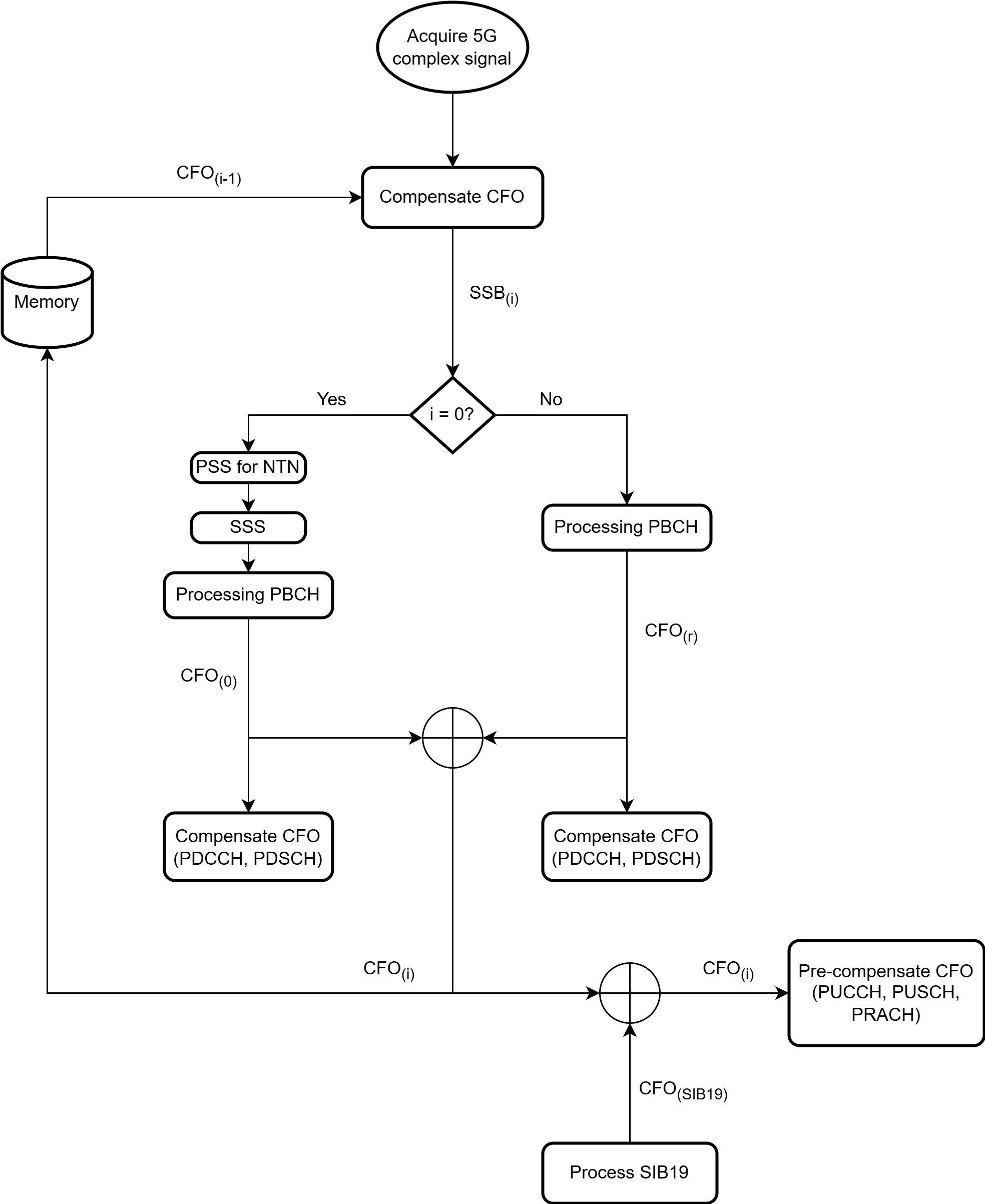}
    \caption{Flowchart of the Doppler compensation process performed by the UE.}
    \label{fig:doppler-compensation-process}
\end{figure}

In the flowchart in Figure \ref{fig:doppler-compensation-process} it is visualized the procedure for DL and UL Doppler compensation.
\section{Field setup}
The LEO satellite has been emulated via a custom Software Defined Radio (SDR) platform mounted on a drone. This platform is equipped with a digital payload that regenerate the signal and applies a specific delay and Doppler shift to the signal arriving and departing to/from the UE, hence effectively emulating the effects given by the movement of a real LEO satellite. The main characteristics of the drone are reported in Table \ref{tab:drone-specs}.
\begin{table}
    \centering
    \caption{Main specifications of the drone.}
    \label{tab:drone-specs}
    \begin{tabular}{c|c}
        \hline
        \textbf{Parameter} & \textbf{Value} \\
        \hline
        Model & QS 17-31 DL/UL \\
        Operating frequency & Ka band (17 -- 31 GHz) \\
        Feeding system & Circular dual-polarized horn \\ 
        \multirow{2}{*}{Polarization} & Dual CP \\
        & LHCP and RHCP \\
        Transmit power (EIRP) & -75 -- +10 dBm \\
        Receiving power & -100 -- +10 dBm \\
        \hline
    \end{tabular}
\end{table}

The UE and gNB are software defined and thus perform only baseband digital processing. For this, they have been equipped with one Universal Software Radio Peripheral (USRP) X300 each that is in charge of performing digital to analog conversion and vice versa. These modems work in the L band, so in order to be able to transmit/receive the signal toward/from the emulated LEO satellite the signal has to pass through a Block UpConverter (BUC) and a Low-Nose Block (LNB) downconverter, respectively, which are responsible for up/down converting the signal frequency to/from the Ka band.

The model of the BUC that has been chosen for the demonstration is the \texttt{SAGE 12W Linear Ka Band GaN}. The main specifications are resumed in Table \ref{tab:buc-specs}. The model of the chosen LNB downconverter instead is the \texttt{Norsat 9000XBF-2}. The main specifications are resumed in Table \ref{tab:lnb-specs}.
\begin{table}
    \centering
    \caption{BUC main characteristics.}
    \label{tab:buc-specs}
    \begin{tabular}{c|c}
        \hline
        \textbf{Parameter} & \textbf{Value} \\
        \hline
        Input frequency & L band (950 -- 1950 MHz) \\
        Output frequency & Ka band (29 -- 31 GHz) \\
        Maximum gain & 55 dB \\
        External reference & 10 MHz \\
        \hline
    \end{tabular}
\end{table}
\begin{table}
    \centering
    \caption{LNB downconverter main characteristics.}
    \label{tab:lnb-specs}
    \begin{tabular}{c|c}
        \hline
        \textbf{Parameter} & \textbf{Value} \\
        \hline
        Input frequency & Ka band (19.20 -- 20.20 GHz) \\
        Output frequency & L band (950 -- 1950 MHz) \\
        Maximum gain & 65 dB \\
        External reference & 10 MHz \\
        Maximum noise figure & 1.6 dBm \\
        \hline
    \end{tabular}
\end{table}

\begin{figure}
    \centering
    \includegraphics[width=\linewidth]{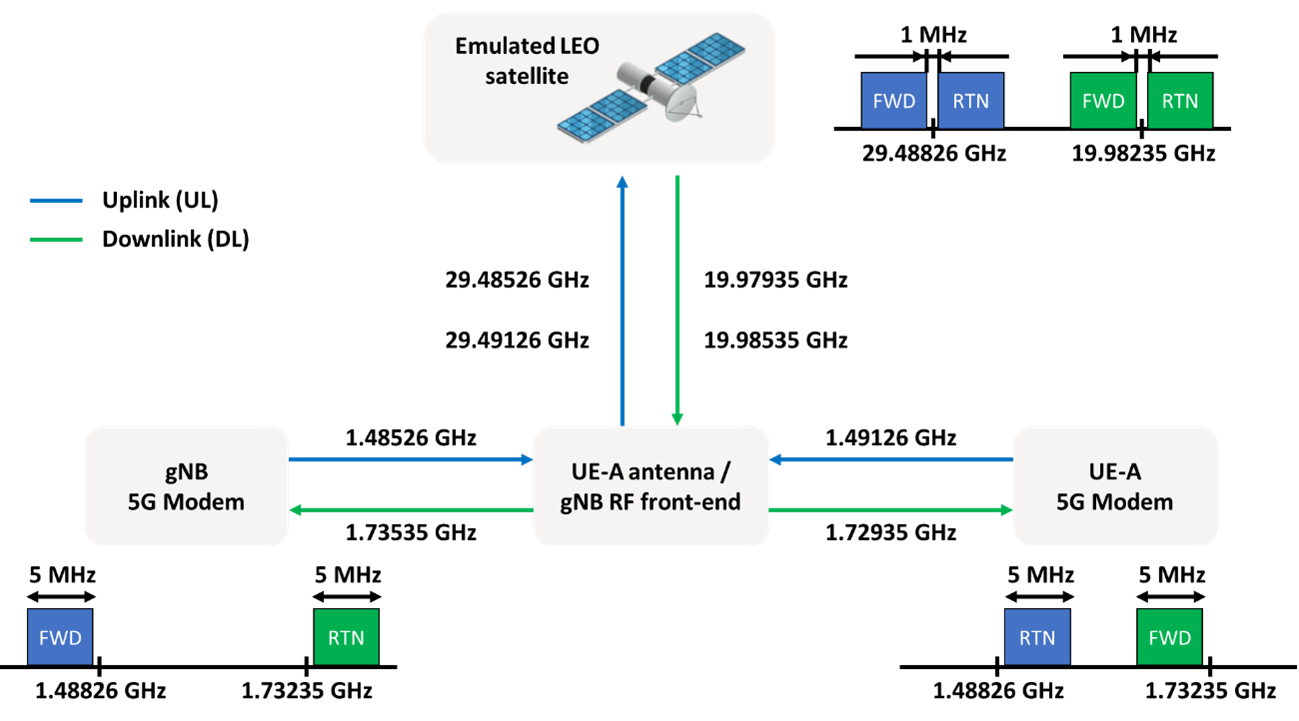}
    \caption{Frequency plan.}
    \label{fig:freq-plan}
\end{figure}

The frequency plan is summarized in Fig.~\ref{fig:freq-plan}. The emulated LEO satellite allocates two slots of 5 MHz -- with 1 MHz guard band -- to the DL and the UL. It also performs a frequency conversion: communication from the gNB to the UE occurs around 29.48826 GHz, whereas in the opposite path it occurs around 19.73826 GHz. The UE and gNB instead operate at a center frequency of 1.48826 GHz.

Summing up the UL and DL slots bandwidth and the guard band, the total bandwidth needed for the system to work is a total of 11 MHz. So, the drone has been configured to sample at a rate of 11 Msps.
\subsection{Wired Test}
\begin{figure}
    \centering
    \includegraphics[width=\linewidth]{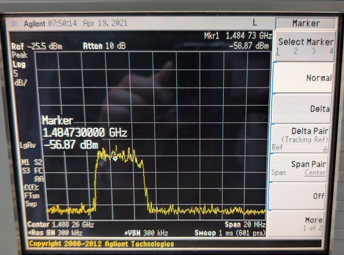}
    \caption{gNB DL signal showing on the spectrum analyzer.}
    \label{fig:spectr-wired1}
\end{figure}

As a first step, we prepared the following setup. The modems of the UE and the gNB were connected through cables to the drone, and the BUC and the LNB downconverter were placed in between such links to perform the L-Ka band conversion. The drone and the LNB downconverter were not introducing any attenuation, while after the BUC we introduced an attenuation of 30 dB to not saturate the payload. Moreover the drone was operating in analog mode, that is, it just relayed the packets in a transparent manner. By analyzing the gNB signal relayed by the drone towards the UE, we measured a Carrier to Noise ratio (C/N) of 15 dB, whereas the signal power had a strength of -57 dBm. In Figure \ref{fig:spectr-wired1} shows the output of the spectrum analyzer.

We repeated the test by enabling the digital processing (regenerative payload) of the drone, with a transmit and receiving attenuation of 30 dB and 10 dB, respectively. This time, we obtained a C/N of 20 dB and a gNB signal strength of -41 dBm, resulting in a enhancement of the operating conditions of the system.

From here, we proceeded to power on the UE to see if the whole 5G system would work. Indeed, the UE successfully managed to acquire the downling synchronization signals advertised by the gNB, completing the UL synchronization procedure (random-access procedure), and finally obtaining an IP from the network.
\subsection{Wireless Test}
\begin{figure}
     \centering
     \begin{subfigure}{\linewidth}
         \centering
         \includegraphics[width=0.75\linewidth]{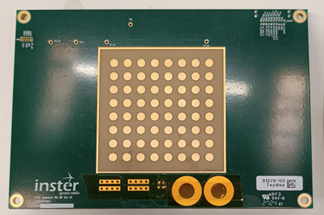}
         \caption{Top layer side.}
     \end{subfigure}
     \hfill
     \begin{subfigure}{\linewidth}
         \centering
         \includegraphics[width=0.75\linewidth]{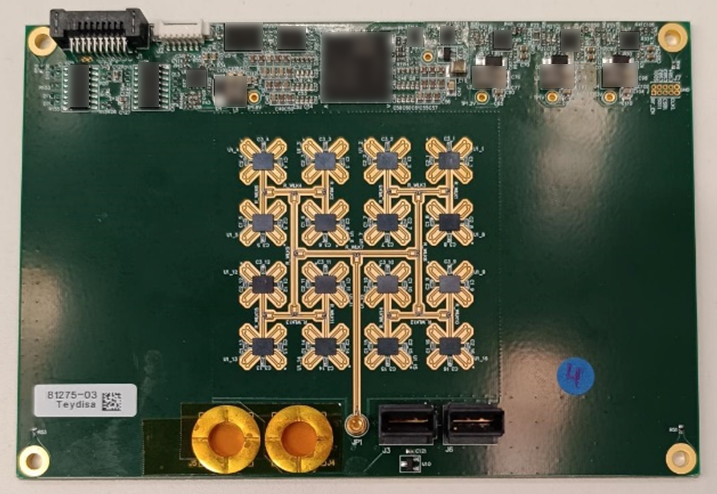}
         \caption{Bottom layer side.}
     \end{subfigure}
     \caption{Receive Ka antenna developed by Inster.}
     \label{fig:rx-ka-ant}
\end{figure}

Once the wired test was successful, we started to prepare the wirleess setup, that consists of two Ka aperture 8x8 electronically steerable antennas developed by Inster, one for transmission and one for reception, which are shared by both the UE and the gNB thanks to the use of signal combiners and splitters. These antennas were implemented on a dedicated Printed Circuit Board (PCB). The top layer of the PCB contains the radiating elements, the bottom layer is dedicated to the beamforming Monolithic Microwave Integrated Circuits (MMICs) and the radio frequency (RF) feeding network, and the internal layer are utilized for the power supply and digital signal routing. Figure \ref{fig:rx-ka-ant} shows the receive antenna. Table \ref{tab:antenna-specs} resumes the specifications of the trasmit and receive antennas.
\begin{table}
    \centering
    \caption{Characteristics of the Ka antennas developed by Inster.}
    \label{tab:antenna-specs}
    \begin{tabular}{c|c}
        \hline
        \textbf{Parameter} & \textbf{Value} \\
        \hline
        \multirow{2}{*}{Operating frequency} & Transmission at 29 -- 31 GHz \\
        & Reception at 19.2 -- 21.2 GHz \\
        Radiating elements & 8x8 microstrip patch antenna \\
        Beamforming & Analog beamforming MMICs with phase shifters \\
        Polarization & Supports electronic polarization switching \\
        \hline
    \end{tabular}
\end{table}

The modems of the UE and the gNB, the BUC and the LNB downconverter share the same 10 MHz reference signal while the drone relies on a 10 MHz reference signal coming from an embedded Global Positioning System Disciplined Oscillator (GPSDO) unit.

\begin{figure}
    \centering
    \includegraphics[width=\linewidth]{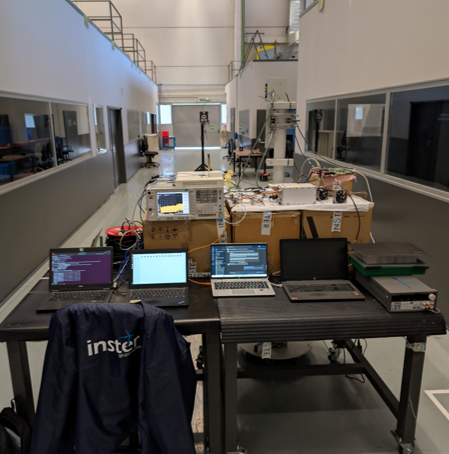}
    \caption{Wireless test setup.}
    \label{fig:wireless-setup}
\end{figure}

A photo of the setup is provided in Figure \ref{fig:wireless-setup}. With this wireless configuration we were successfully able to close the physical link, with received signal strength and C/N aligning with the link budget estimations. This outcome confirms the viability of the RF and frequency translation chain. A comprehensive end-to-end system evaluation will follow, as outlined in the conclusion.

\section{Conclusion and Future Work}
This paper presented the design of a 5G UE adapted to operate in LEO-based NTN scenarios, with specific focus on handling variable delay and high Doppler shifts. We detailed the compensation strategies implemented at the UE level and described the architecture of a custom field testbed built to emulate LEO satellite conditions using a drone-mounted SDR platform. While full end-to-end system validation is still underway, initial lab tests -- including link budget analysis and spectrum inspection -- have confirmed the feasibility and proper configuration of the signal chain. In future work, we plan to finalize the testbed setup and carry out complete system-level evaluations under both transparent and regenerative payload modes. Performance evaluations and experimental results will be presented in future work once the testbed is fully operational.
\printbibliography
\end{document}